# CRYOPRESERVATION OF SUGARCANE EMBRYOGENIC CALLUS USING A SIMPLIFIED FREEZING PROCESS

M.E. Martínez-Montero[1]*, M.T. González-Arnao[2], C. Borroto-Nordelo[1], C. Puentes-Díaz[1] and F. Engelmann[3]

1: Centro de Bioplantas, UNICA, Car. a Morón km 9, CP-69450, Ciego de Ávila, Cuba.
2: Centro Nacional de Investigaciones Científicas (CNIC), Ave. 25 y 158, Apartado 6990, Cubanacán, Playa, La Habana, Cuba.
3: International Plant Resources Institute (IPGRI), Via delle Sette Chiese 142, 00145 Rome, Italy.

**Summary:** A simplified freezing process was developed and successfully applied to embryogenic calluses of three sugarcane commercial hybrids (*Saccharum* sp. cv. CP 5243, C 91-301 and C 1051-73). To obtain optimal survival, the calluses were pretreated with a cryoprotective solution containing 10% DMSO and 0.3 to 0.75M sucrose. For freezing, the samples were immersed in an alcohol bath placed in a -40°C freezer, thus allowing a freezing rate comprised between 0.4 and 0.6°C/min. Samples were held at that temperature for 2 h before immersion in liquid nitrogen. The highest survival rates of cryopreserved calluses ranged between 20 and 94% depending on the variety, and fully developed plantlets could be obtained from regenerating calluses. Embryogenic calluses of one variety were stored for 14 months in liquid nitrogen without any effect on their survival rate and plantlet production.

## INTRODUCTION

In addition to its application for the conservation of plant genetic resources, cryopreservation is also a useful tool for the management and conservation of material produced in vitro. Storing embryogenic cell suspensions or calluses in liquid nitrogen preserves their regeneration capacity and limits the risks of somaclonal variation which increase with culture duration.

Classical freezing protocols, which include slow controlled cooling (0.5 - 1°C.min$^{-1}$) down to around -40°C followed by immersion of samples in liquid nitrogen, are usually employed for cell suspensions and embryogenic calluses (6). Classical cryopreservation protocols have been developed for embryogenic calluses of numerous species. Simplified freezing protocols in which the expensive programmable freezers required for classical protocols are replaced with simpler freezing devices including alcohol baths and/or domestic freezers have been established for various materials (3, 4, 12, 14, 16, 17, 19).

In the case of sugarcane, cryopreservation protocols have been developed for various materials: apices of *in vitro* plantlets have been frozen using the encapsulation-dehydration technique (8, 9, 10, 15); cell suspensions (1, 5) and embryogenic calluses have been cryopreserved using classical freezing protocols (2, 7, 11, 18).

In this paper, we describe a simplified cryopreservation protocol and its application to embryogenic callus of three commercial varieties of sugarcane. We also present data on the effect of extended cryopreserved storage duration on the survival and plantlet production of calluses of one sugarcane variety.

## MATERIALS AND METHODS
### *In vitro* culture

Embryogenic calluses were initiated by culturing 3-5 mm long immature inflorescence segments on MS (13) medium supplemented with 3.0 mg.L$^{-1}$ 2,4-dichlorophenoxyacetic acid (2,4-D) for 30 d. Calluses were subcultured monthly on MS medium supplemented with 2,4-D (1.0 mg.L$^{-1}$), arginine (50 mg.L$^{-1}$) and proline (500 mg.L$^{-1}$). They were cultured in the dark at 25 ± 2C. For plantlet regeneration, calluses were transferred to a medium devoid of 2,4-D at 27 ± 2°C under a 16 h light/8 h dark photoperiod, with a 40 µmol.m$^{-2}$s$^{-1}$ photon dose.

### Cryopreservation

For cryopreservation experiments, 15 to 25 day-old calluses, about 3 to 5 mm in diameter were employed. About one-third of sterile 2 ml cryotubes were filled with fragments of calluses which were pretreated in liquid medium with sucrose concentrations ranging between 0.3 and 0.75M for 1 h at 0°C. Dimethylsulfoxide (DMSO) was added progressively to the liquid medium over a period of 30 min until the final concentration (5, 10 or 15%, v/v) was reached.

Freezing was performed using a home-made ethanol bath, consisting of a polypropylene container filled with 700 ml ethanol precooled at 0°C. Cryotubes were inserted in holes pierced in a thin polypropylene plate floating on top of the ethanol, which allowed immersion of the cryotubes in the coolant. The ethanol bath was then placed in a -40°C freezer, thus allowing an average cooling rate of 0.4-0.6°C.min$^{-1}$ between 0°C and -40°C. The cooling rate of the system was recorded during all experiments using a digital thermometer with a copper-constantan thermocouple immersed in 1 ml of the cryoprotective solution contained in a cryotube, which was frozen as described previously. Crystallization was induced manually in the cryoprotective medium at a temperature intermediate between the nucleation and the crystallization temperature of the cryoprotective medium, by briefly putting the base of the cryotubes in contact with liquid nitrogen. The temperature at which crystallization of the cryoprotective medium was induced ranged between -7°C and -20°C. Once the temperature of -40°C was reached, the cryotubes were kept for 2 h at this temperature, then immersed rapidly in liquid nitrogen. Samples were kept for a minimum of 2 h at -196°C. Rapid thawing was carried out by plunging the cryotubes in a +40°C water-bath. Calluses were then transferred directly (without washing) to recovery medium (MS medium supplemented with 2,4-D (1.0 mg.L$^{-1}$), arginine (50 mg.L$^{-1}$) and proline (500 mg.L$^{-1}$)).

Three replicate cryovials, each containing six fragments of embryogenic calluses were used for each experimental condition. The survival rate, evaluated 40 to 50 d after freezing, corresponded to the percentage of calluses which had increased in size during the recovery period. For the measurement of plantlet production, six calluses randomly chosen in each condition were transferred to standard medium without 2,4-D, and the number of plantlets produced was estimated after 80 d of culture. The survival rate and plantlet production of calluses of variety CP 5243 were evaluated after storage for 2 h, 4 and 14 months in liquid nitrogen.

**RESULTS**

Survival of calluses after pretreatment was comparable between the three varieties (Table l). It was high with 5 and 10% DMSO, and decreased slightly with 15% DMSO. After freezing in liquid nitrogen, higher survival was achieved with variety CP 5243 than with the other two varieties. With varieties CP 5243 and C 91-301 high survival rates were obtained for a large range of pretreatment conditions. In contrast, for variety C 1051-73, high survival was achieved after pretreatment with 5% DMSO and 0.3 M sucrose only.

**Table 1:** Effect of DMSO and sucrose concentration in the cryoprotective solution on the survival rate (%) of control (-LN) and cryopreserved (+LN) embryogenic calluses of sugarcane varieties CP 5243, C 91-301 and C 1051-73.

| DMSO (%) | Sucrose (M) | Survival | | | | | |
|---|---|---|---|---|---|---|---|
| | | CP 5243 | | C 91-301 | | C 1051-73 | |
| | | -LN | +LN | -LN | +LN | -LN | +LN |
| 5 | 0.3 | 100 | 82 | 94 | 0 | 94 | 0 |
| | 0.5 | 100 | 82 | 93 | 6 | 89 | 8 |
| | 0.75 | 100 | 85 | 94 | 9 | 89 | 6 |
| 10 | 0.3 | 97 | 81 | 94 | 30 | 88 | 41 |
| | 0.5 | 100 | 90 | 94 | 22 | 87 | 17 |
| | 0.75 | 100 | 94 | 83 | 22 | 81 | 14 |
| 15 | 0.3 | 84 | 61 | 84 | 20 | 78 | 6 |
| | 0.5 | 86 | 58 | 78 | 25 | 76 | 6 |
| | 0.75 | 83 | 60 | 78 | 20 | 67 | 21 |

Fully developed plants could be obtained from regeneratmg calluses of all three sugarcane varieties (Table 2). The number of plantlets produced from control calluses was higher than from cryopreserved ones and regeneration from calluses of variety CP 5243 was much higher than from the two other varieties. Regeneration of plantlets was obtained from a much broader range of experimental conditions than those which had ensured optimal survival.

**Table 2:** Effect of DMSO and sucrose concentration in the cryoprotective solution on the number of plantlets produced from control (-LN) and cryopreserved (+LN) embryogenic calluses of sugarcane varieties CP 5243, C 91-301 and C 1051-73.

|  |  | Plantlets produced | | | | | |
|---|---|---|---|---|---|---|---|
|  |  | CP 5243 | | C 91-301 | | C 1051-73 | |
| DMSO (%) | Sucrose (M) | -LN | +LN | -LN | +LN | -LN | +LN |
| 5 |  | 190 | 143 | 88 | 0 | 90 | 0 |
|  |  | 196 | 165 | 69 | 33 | 90 | 35 |
|  |  | 211 | 150 | 72 | 34 | 88 | 44 |
| 10 |  | 210 | 150 | 82 | 40 | 95 | 48 |
|  |  | 213 | 106 | 73 | 42 | 99 | 51 |
|  |  | 200 | 119 | 63 | 41 | 88 | 45 |
| 15 |  | 243 | 165 | 70 | 18 | 82 | 27 |
|  |  | 209 | 171 | 65 | 18 | 81 | 29 |
|  |  | 209 | 161 | 51 | 18 | 75 | 25 |

The effect of storage duration on the survival rate and plantlet regeneration of cryopreserved calluses of variety CP 5243 is presented in Table 3. Even though a slight decrease in survival between 2 h and 4 months in storage could possibly be noted under some conditions, both the survival rate and the number of plantlets regenerated from cryopreserved calluses were not modified after 14 months under LN storage.

**Table 3:** Effect of extended storage duration on the survival rate and plantlet production of callus of variety CP 5243.

|  |  | Survival (%) | | | Plantlets produced | | |
|---|---|---|---|---|---|---|---|
| DMSO (%) | Sucrose (M) | 2 h | 4 months | 14 months | 2 h | 4 months | 14 months |
| 10 | 0.3 | 81 | 75 | 77 | 150 | 113 | 95 |
|  | 0.5 | 90 | 81 | 78 | 106 | 124 | 115 |
|  | 0.75 | 94 | 73 | 77 | 119 | 99 | 100 |

## DISCUSSION/CONCLUSION

The simplified freezing protocol developed in this study was deemed efficient since it achieved results comparable to those obtained with the classical protocols employed by other authors to cryopreserve sugarcane calluses (2, 7, 11).

During preculture experiments, DMSO was the cryoprotectant determining the toxicity of the various cryoprotective mixtures tested. Differences in sensitivity to the different cryoprotective mixtures were noted between clones, as previously reported by Eksomtramage *et al*. (2).

Previous experiments (2, 11) using a classical freezing procedure had indicated that sugarcane embryogenic calluses could be successfully frozen using a relatively wide range of cooling rates, between 0.1 to 1°C.min$^{-1}$. This allowed us to obtain positive results after freezing the embryogenic calluses with a cooling rate ranging between 0.4 and 0.6°C.min$^{-1}$.

In this work, the use of a simplified freezing protocol achieved good survival and numerous plantlets from regenerating calluses of all sugarcane varieties. The variety CP 5243 gave a much higher survival rate and production of plantlets than the other two varieties tested. However, it is important to mention that control calluses of these two varieties (C 91-301 and C 1051-73) had a much slower growth rate and released a large amount of phenolic compounds in the medium. This underlines the importance of the in vitro propagation procedures in the successful establishment of a cryopreservation protocol for any given material.

Extending storage duration in liquid nitrogen to 14 months did not induce any modification in the survival rate and plantlet production of cryopreserved calluses, which confirms the applicability of cryopreservation for the long-term conservation of material produced *in vitro*.

In conclusion, efficient and simple cryopreservation protocols have been developed fop cell suspensions, embryogenic calluses and apices of sugarcane and successfully applied to a wide range of genotypes. Sugarcane should thus be one of the first crops to receive the routine application of cryopreservation.

## ACKNOWLEDGEMENTS
The authors gratefully thank IPGRI for allowing F.E. to participate in this research project.